# To Optimally Design Microstrip Nonuniform Transmission Lines as Lowpass Filters

M. Khalaj-Amirhosseini and S. A. Akbarzadeh-Jahromi

**Abstract**— A method is proposed to optimally design the Microstrip Nonuniform Transmission Line (MNTLs) as lowpass filters. Some electrical and physical restrictions are used to design MNTLs. To optimally design the MNTLs, their strip width is expanded as truncated Fourier series, firstly. Then, the optimum values of the coefficients of the series are obtained through an optimization approach. The performance of the proposed structure is studied by design and fabrication of two lowpass filters of cutoff frequency 2.0 GHz.

**Index Terms**—Nonuniform Transmission Lines, Lowpass Filters, Microstrip Lines.

———————————— ◆ ————————————

## 1 INTRODUCTION

A commonly used structure for microstrip lowpass filters (LPFs) is Hi-Z, Lo-Z or stepped-impedance structure in which several very high and very low impedance transmission lines are located alternately [1-3]. However the performance of these popular filters is not as good due to the approximations involved in their design. Moreover, the stepped-impedance filters have discontinuities, which create some implementation constraints. On the other hand, Microstrip Nonuniform Transmission Lines (MNTLs) are widely used in microwave circuits as resonators [4], impedance matching [4]-[5], delay equalizers [6], wave shaping [7], analog signal processing [8] etc. Some researchers have used MNTLs to design lowpass or bandreject filters with specific response functions such as Butterworth or Chebyshev types [9-11]. The MNTL based lowpass filters have not any discontinuity and quasi-TEM approximation is more valid for them than to stepped-impedance filters. This paper presents a method to design optimally the MNTLs as lowpass filters. In the presented method, some arbitrary restrictions are used instead of a known specific response. Moreover, the strip width of MNTLs is considered as a truncated Fourier series, instead of a collection of several uniform or cubic sections [10]. The optimum values of the coefficients of the Fourier series are obtained through an optimization approach. Finally, the performance of the proposed structure is studied using two experiments.

## 2 ANALYSIS OF MNTL LPFs

In this section MNTLs are considered as lowpass filters. Fig. 1 depicts a typical Microstrip Nonuniform Transmission Line (MNTL) of length $d$ terminated by source and load resistances $Z_0$. The relative electric permittivity and the thickness of the substrate are $\varepsilon_r$ and $h$, respectively. Also, the width of strip is varying with $z$ as $w(z)$. We would like to design MNTLs as lowpass filters specified by arbitrary restrictions shown in Fig. 2. The passband and stopbands are defined in the frequency ranges $[0 - f_P]$ and $[f_s - f_{max}]$, respectively. The frequency response of designed filter must be kept out of hachuring regions of Fig. 2.

To analyze MNTLs, one can use some quasi-TEM based methods such as cascading many short sections [12, 13], finite difference [14], Taylor's series expansion [15], Fourier series expansion [16], the equivalent sources method [17], the method of Moments [18] and some approximated closed form solutions [19]. Of course, the most straightforward method is subdividing MNTLs into many uniform or linear electrically short sections of length $\Delta z$ so that

$$\Delta z << \lambda_{min} = \frac{c}{f_{max}\sqrt{\varepsilon_r}} \quad (1)$$

where $c$ is the velocity of the light. Then the *ABCD* matrix of the MNTL will be obtained by multiplying the *ABCD* matrices of all short sections. After finding the *ABCD* parameters, one can determine the *S* parameters as follows

$$S_{11} = \frac{AZ_0 + B - CZ_0^2 - DZ_0}{AZ_0 + B + CZ_0^2 + DZ_0} \quad (2)$$

$$S_{21} = \frac{2Z_0}{AZ_0 + B + CZ_0^2 + DZ_0} \quad (3)$$

## 3 SYNTHESIS OF MNTL LPFs

In this section a general method is proposed to optimally design MNTLs as lowpass filters. First, we consider the following truncated Fourier series expansion for the normalized width function $w(z)/h$.

————————————————

- *M. Khalaj-Amirhosseini is with Iran University of Science and Technology, Tehran, Iran.*
- *S. A. Akbarzadeh-Jahromi is with Iran University of Science and Technology.*

$$\ln\left(\frac{w(z)}{h}\right) = \sum_{n=0}^{N} C_n \cos(2\pi n z/d) + \sum_{n=1}^{N} S_n \sin(2\pi n z/d) \quad (4)$$

The optimum values of the unknown coefficients $C_n$ and $S_n$ in (11) can be obtained through minimizing the following defined error function.

$$\text{Error} = \sqrt{\frac{1}{N_f}\left(\sum_{0<f\le f_p}|S_{11}(f)|^2 + \sum_{f_p<f\le f_{max}}|S_{21}(f)|^2\right)} \quad (5)$$

where $N_f$ is total number of frequencies in the range of zero to $f_{max}$. Moreover, the above defined error function should be restricted by some electrical (as are seen in Fig. 2) and physical constraints like as the followings

$$\min(20\log(|S_{21}(f)|)) \ge -\alpha_p \ ; \ 0 < f \le f_p \quad (6)$$

$$\max(20\log(|S_{21}(f)|)) \le -\alpha_s \ ; \ f_s \le f \le f_{max} \quad (7)$$

$$20\log(|S_{21}(f)|) \le -\alpha_p - \frac{\alpha_s - \alpha_p}{f_s - f_p}(f - f_p) \ ; \ f_p < f < f_s \quad (8)$$

$$\left(\frac{w}{h}\right)_{min} \le \frac{w(z)}{h} \le \left(\frac{w}{h}\right)_{max} \quad (9)$$

$$\frac{w(0)}{h} = \frac{w(d)}{h} = \frac{w_0}{h} \quad (10)$$

where $(w/h)_{min}$ and $(w/h)_{max}$ are the minimum and maximum available normalized width, respectively. Also, $w_0/h$ is desired normalized width at the ends of the microstrip lines corresponding to desired characteristic impedance $Z_0$.

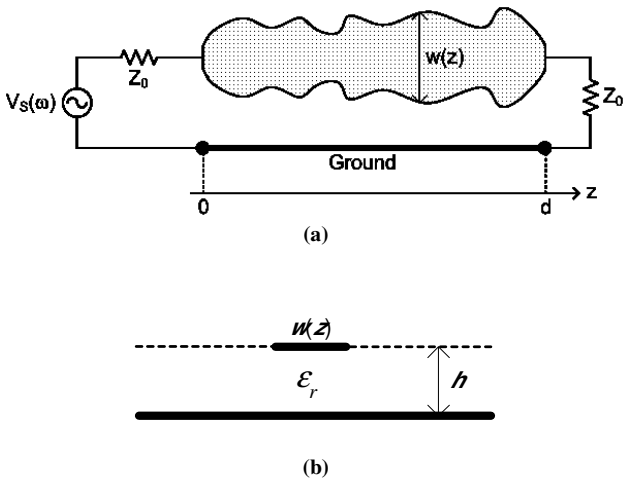

**Figure 1. Typical Microstrip Nonuniform Transmission Line
a) Longitudinal view  b) The cross section**

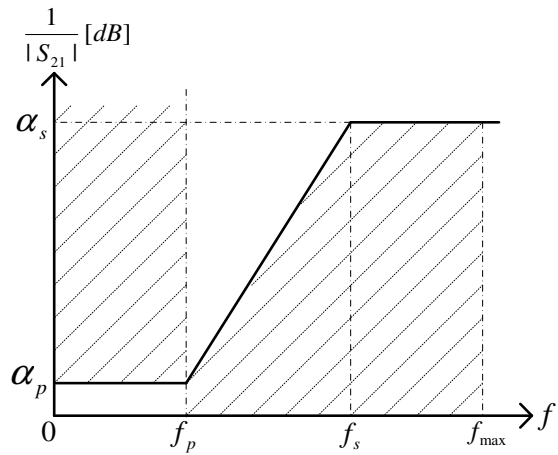

**Figure 2. Defined restrictions for a lowpass filter (out of hachuring regions)**

## 4 EXAMPLES AND RESULTS

In this section two MNTL lowpass filters are designed and fabricated. The specifications of these filters are considered as the following:

1. LPF No. 1: $f_p$ = 2.0 GHz, $f_s$ = 3.0 GHz, $f_{max}$ = 6.0 GHz, $\alpha_p$ = 0.1 dB, $\alpha_s$ = 20 dB, $(w/h)_{max}$ = 10, $(w/h)_{min}$ = 0.13, $d$ = 10.0 cm and $Z_0$ = 50 Ω.

2. LPF No. 2: $f_p$ = 2.0 GHz, $f_s$ = 3.0 GHz, $f_{max}$ = 6.0 GHz, $\alpha_p$ = 0.3 dB, $\alpha_s$ = 20 dB, $(w/h)_{max}$ = 7, $(w/h)_{min}$ = 0.1, $d$ = 10.0 cm and $Z_0$ = 50 Ω.

The filters were designed and simulated by HFSS full-wave software and then were fabricated on a substrate with $\varepsilon_r$ = 3.5 and $h$ = 30 mil = 768 μm. Figs. 3-5 show the normalized width functions and the pictures of the fabricated filters. Also, Table 1 shows the optimum values of the unknown coefficients $C_n$ and $S_n$ for designed filters. In continuation, Figs. 6 and 7 compare the response of the filters obtained from analysis, simulation and measurement. It is seen that the agreement between theoretical and measurement results is good, especially for filter No. 2 whose allowable attenuation in the passband is more than that of filter No. 1. The extra losses in the passband may be due to the losses of substrate and connectors, not considering the thickness of conductive strip, radiation and weak validation of the quasi-TEM approximation at wide regions. It is expected as the length of the filter is chosen larger the required width of conductive strip is decreased and consequently the quasi-TEM approximation becomes more validated.

**Table 1. Optimum values of the coefficients $C_n$ and $S_n$**

| LPF | $C_0$ | $C_1$ | $C_2$ | $C_3$ | $C_4$ | $C_5$ |
|---|---|---|---|---|---|---|
| No. 1 | 0.3805 | 0.2716 | −0.0143 | −0.1071 | −0.4725 | 0.7393 |
| No. 2 | 0.2333 | 0.3900 | −0.0637 | −0.0078 | −0.6005 | 0.8461 |
| | − | $S_1$ | $S_2$ | $S_3$ | $S_4$ | $S_5$ |
| No. 1 | − | −0.1593 | −0.0968 | −0.1729 | −0.8906 | 1.1364 |
| No. 2 | − | −0.2200 | 0.0929 | 0.0569 | −1.0636 | 0.5341 |

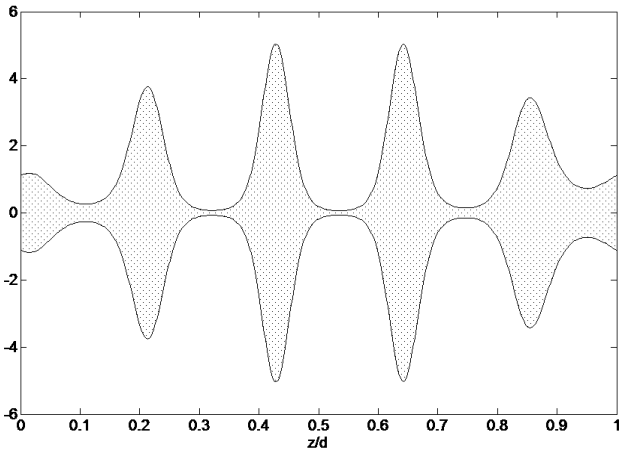

Figure 3. The normalized width function *w*(*z*)/*h* of filter No. 1

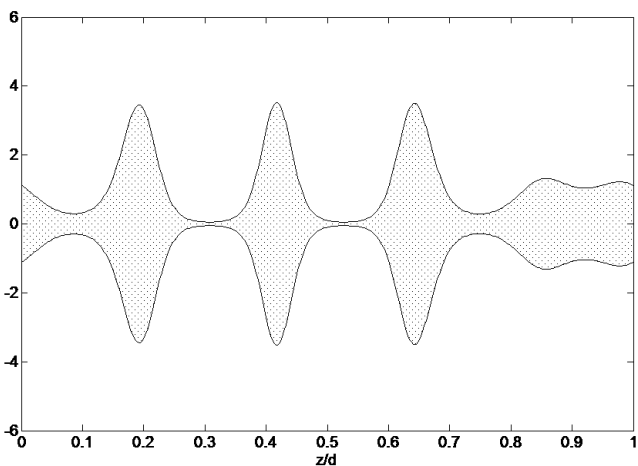

Figure 4. The normalized width function *w*(*z*)/*h* of filter No. 2

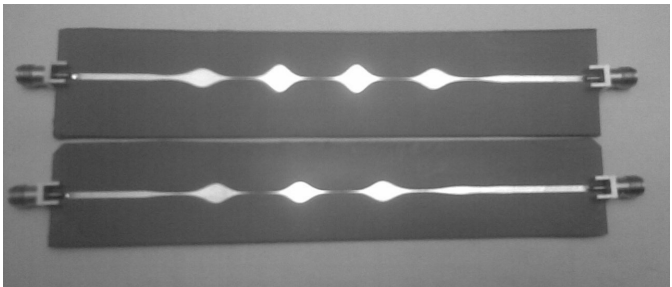

Figure 5. The picture of fabricated filters   a) filter No. 1   b) filter No. 2

## 5 CONCLUSION

A method was proposed to optimally design the Microstrip Nonuniform Transmission Line (MNTLs) as lowpass filters. Some arbitrary electrical and physical restrictions are used to design MNTLs. To optimally design the MNTLs, their strip width is expanded as truncated Fourier series, firstly. Then, the optimum values of the coefficients of the series are obtained through an optimization approach. Two lowpass filters of cutoff frequency 2.0 GHz were designed, fabricated and measured. The agreement between theoretical and measurement results was satisfactory, especially for filters whose allowable attenuation in the passband is high. The MNTL based lowpass filters have not any discontinuity. Finally, it is worthy to mention that the proposed method can be used to design MNTLs as Bandreject filters by suitable defining the electrical restrictions.

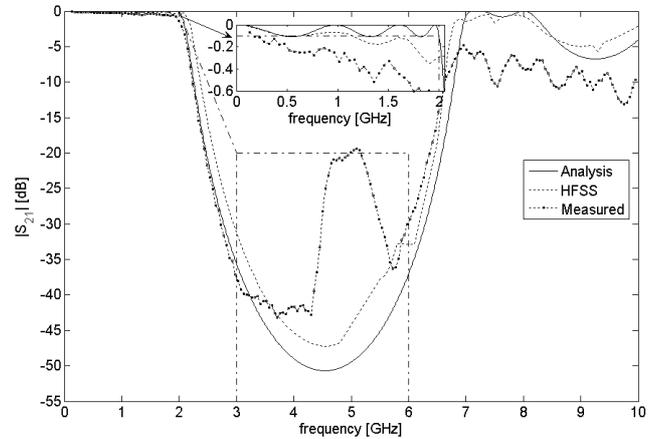

Figure 6. The absolute of $S_{21}$ for filter No. 1

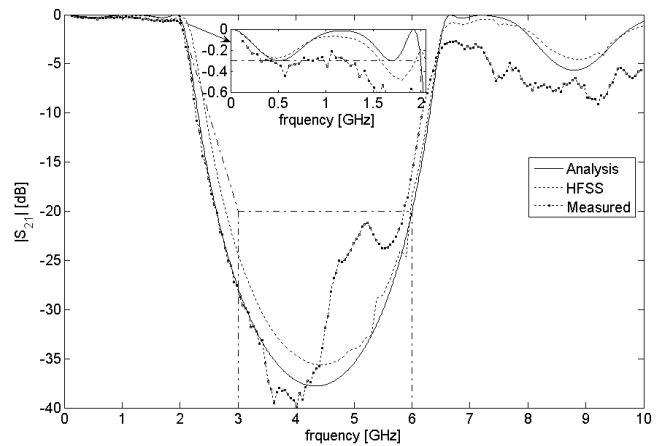

Figure 7. The absolute of $S_{21}$ for filter No. 2

**Mohammad Khalaj Amirhosseini** was born in Tehran, Iran in 1969. He received his B.Sc, M.Sc and Ph.D. degrees from Iran University of Science and Technology (IUST) in 1992, 1994 and 1998 respectively, all in Electrical Engineering. He is currently an Associate Professor at College of Electrical Engineering of IUST. His scientific fields of interest are electromagnetic direct and inverse problems including microwaves, antennas and electromagnetic compatibility.

**Abbas Akbarzadeh** was born in Jahrom, Iran, on April 13, 1984. He received the B.S. and M.S. degrees (with honors) in electrical engineering from the University of Iran University of Science and Technology "IUST", Tehran, Iran, in 2007. His research interests include of RF/microwave passive structures and antennas, and circuit components in presence of complex materials.